\documentclass[aps,prl,twocolumn,showpacs,groupedaddress,amsmath,amssymb,floatfix,tightenlines]{revtex4-1}
\usepackage{graphicx}
\usepackage{graphics}
\graphicspath{{Figures/}}

\begin{document}



\title{Strong orientational coupling of block copolymer microdomains to smectic layering revealed by magnetic field alignment}

\author{Manesh Gopinadhan}
\email[]{manesh.gopinadhan@yale.edu}
\affiliation{Department of Chemical Engineering, Yale University, New Haven CT 06511}
\author{Youngwoo Choo}
\email[]{youngwoo.choo@yale.edu}
\affiliation{Department of Chemical Engineering, Yale University, New Haven CT 06511}
\author{Chinedum O. Osuji}
\email[]{chinedum.osuji@yale.edu}
\affiliation{Department of Chemical Engineering, Yale University, New Haven CT 06511}


\date{\today}

\begin{abstract}
We elucidate the roles of the isotropic-nematic (I-N) and nematic-smectic A (N-SmA) transitions in magnetic field directed self-assembly of a liquid crystalline block copolymer (BCP), using \textit{in situ} x-ray scattering. Cooling into the nematic from the disordered melt yields poorly ordered and weakly aligned BCP microdomains. Continued cooling into the SmA however results in an abrupt increase in BCP orientational order with microdomain alignment tightly coupled to the translational order parameter of the smectic layers. These results underscore the significance of the N-SmA transition in generating highly aligned states under magnetic fields in these hierarchically ordered materials.
\end{abstract}


\maketitle


Block copolymers (BCPs) are versatile materials that are valued for their combination of tunable mesoscopic structural length scales and diverse chemical functionality. BCP self-assembly is well-understood for the canonical case of a diblock system where both blocks are Gaussian coils. A generic sequence of spheres, cylinders and lamellae is encountered as a function of volume fraction at constant temperature \cite{matsen1996unifying,bates2008block}. The order-disorder temperature, $\mathrm{T_{odt}}$, delineates the self-assembled or ordered system from a typically high temperature disordered state. There is considerable interest in directed self-assembly (DSA) as a means to control orientational and positional order in BCP mesophases. Such control is critical for leveraging the useful length scales and chemical functionality of BCPs in scenarios ranging from advanced lithography to membrane separations \cite{park2003enabling,darling2007dsa,hu2014directed}.

A common strategy for DSA involves the use of external fields. The ease with which magnetic fields can be applied in arbitrary geometries and without risk of dielectric breakdown represent beneficial departures from shear and electric field DSA \cite{majewski2012magnetic}. However, the need for substantive magnetic anisotropy often requires the incorporation of rigid anisotropic moieties in the BCP. This has been accomplished in liquid crystalline (LC) BCPs, hierarchically ordered systems wherein mesogens attached to the polymer backbone form LC phases within the BCP superstructure. The magnetic anisotropy of the LC combined with the mesogen anchoring condition at the block interface determine the equilibrium state under the field. While this strategy has been successfully advanced in side-chain LC BCPs \cite{Osuji_Macromol2004,Hamley2004,Xu_magnetic_align_FaradayDisc2009,gopinadhan2012magnetic,gopinadhan2013order,deshmukh2014molecular} and rod-coil systems \cite{Segalman_magnetic_2007}, there is little known regarding the interplay of the coexisting LC and BCP mesophases during alignment. Efforts to elucidate this interplay are well motivated by a desire to better understand the generic features of hierarchical self-assembly in soft matter systems and self-assembly under external fields. Additionally, such efforts fit into a broader narrative regarding DSA as a route to fabricating functional nanostructured materials.

\begin{figure}
\centering
\includegraphics[width=80mm, scale=1]{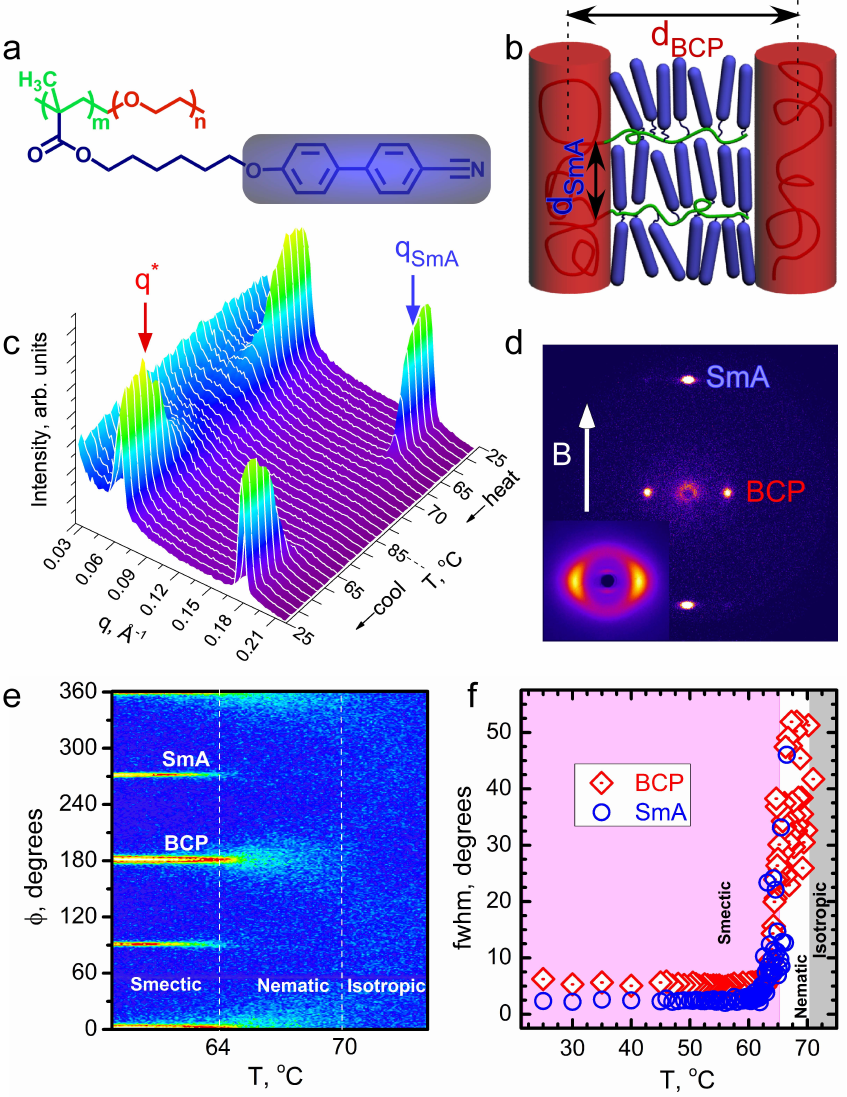}
\caption{a) Structure of LC BCP. b) Hierarchical self-assembly of hexagonally packed PEO cylindrical microdomains within a SmA matrix. c) Temperature resolved SAXS of the LC BCP. d) 2-D SAXS and WAXS (inset) from 6 T field aligned material. e) Azimuthal intensity distribution during slow cooling at 6 T from isotropic melt, and f) corresponding FWHM of BCP and SmA scattering.}
\label{material_system}
\end{figure}

Here, we examine the roles of the isotropic-nematic (I-N) and nematic-smectic A (N-SmA) transitions of an LC mesophase on the field alignment of the BCP superstructure using \emph{in situ} small angle x-ray scattering (SAXS). We conduct zero-field cooling experiments in which the system is exposed to the field during the I-N transition, then cooled through the N-SmA at zero field. A systematic series of temperature dependent data reveals pronounced coupling of the BCP orientational order with that of the smectic layers formed at the N-SmA transition, and concomitant increases of the correlation lengths of both structures. The emergence of translational order in the SmA imposes an increase in physical persistence of the cylindrical BCP microdomains which in turn couples the orientational order of the BCP more strongly to that of the LC director field. This hypothesis is supported by temperature resolved TEM which shows structural evolution in a narrow temperature range from poorly-ordered, low-aspect ratio microdomains in the nematic state, to microdomains with greater physical persistence and lateral order in the smectic.

\begin{figure}
\centering
\includegraphics[width=70mm, scale=1]{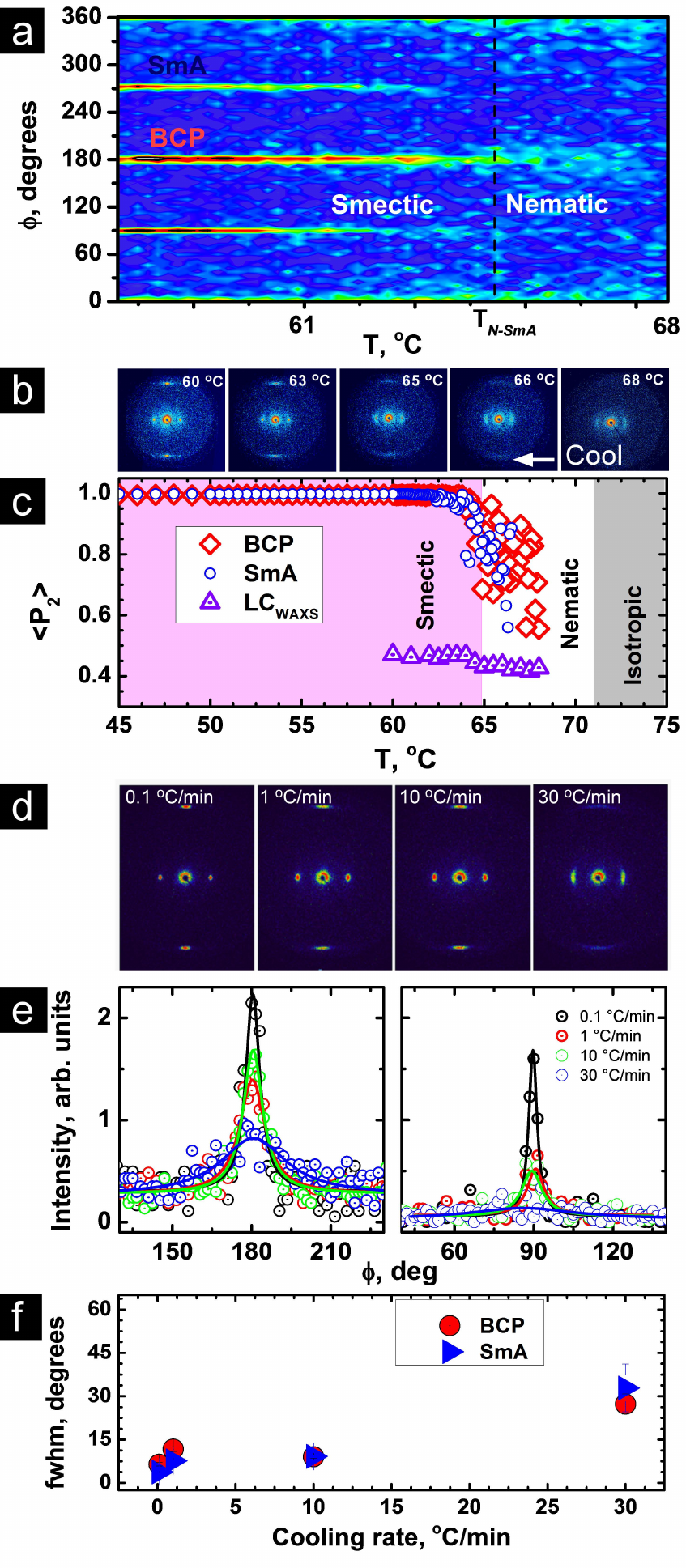}
\caption{a) Azimuthal intensity map and (b) selected SAXS patterns, during zero-field cooling through N-SmA. c) Temperature dependent orientation distribution coefficients, $\langle P_2\rangle$, show strong coupling of BCP alignment with smectic layer orientation. d) SAXS patterns at 62 $^{\circ}$C at different zero-field cooling rates indicated. e) Azimuthal dependence of scattered intensity for SmA layers (left) and BCP domains (right). f) FHWM of azimuthal intensity for different cooling rates.}
\label{alignment_data}
\end{figure}

\begin{figure}
\centering
\includegraphics[width=80mm, scale=1]{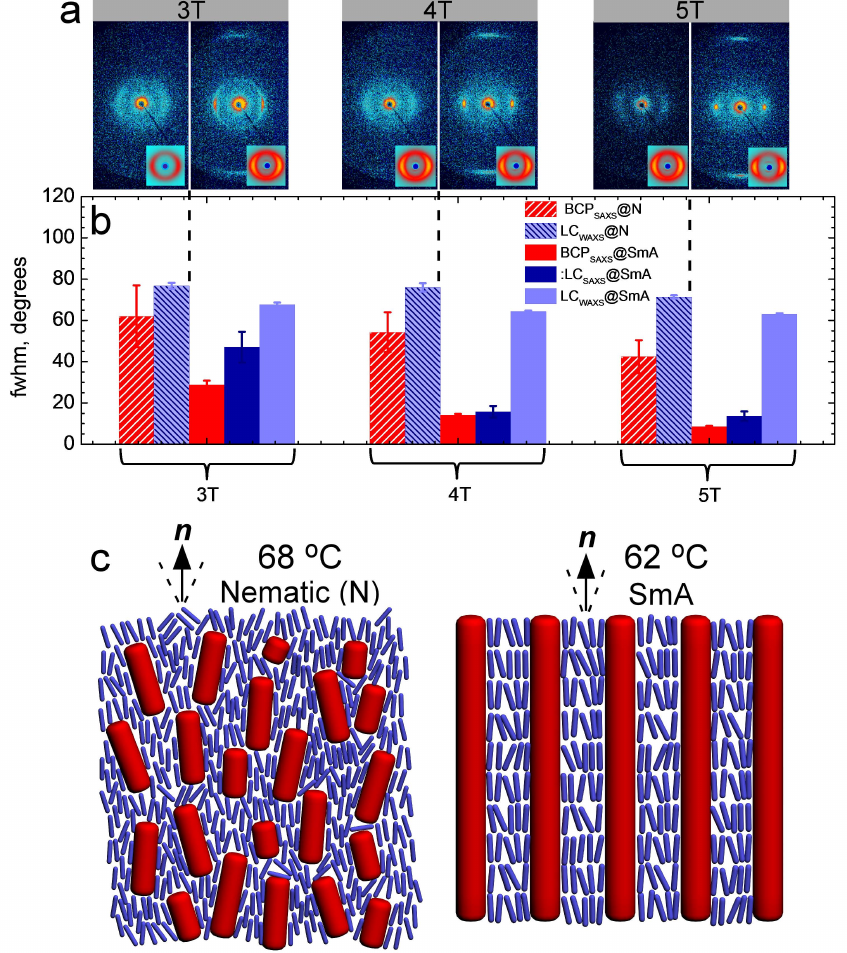}
\caption{a) Scattering patterns (WAXS inset) at different field strengths as indicated and corresponding azimuthal intensity peak widths of BCP and LC structures (SmA layers from SAXS; mesogen packing from WAXS) in N (left) and SmA (right) phases. b) The data indicate a coupling of BCP alignment to smectic layer orientation rather than mesogen orientation, and that the increase in mesogen order from N to SmA is moderate. c) Schematic showing the structural organization of the material in N and SmA states.}
\label{alignment_data_cartoon}
\end{figure}

The system is a side-chain LC BCP, poly(ethylene oxide-b-methacrylate), PEO-b-PMA/LC, in which the PMA block is side-functionalized by cyanobiphenyl mesogens, Fig. \ref{material_system}a. Materials were obtained from PolymerSource. The molecular weight is 10.4 kg/mol. with PEO weight fraction $\mathrm{f_{PEO}=0.23}$. At room temperature the system forms hexagonally packed cylindrical PEO microdomains spaced by $\mathrm{d_{BCP}}$=11 nm embedded in a SmA matrix with a layer spacing $\mathrm{d_{SmA}}$=3.5 nm. The mesogens exhibit planar anchoring at the block interface such that the LC director is parallel to the cylindrical microdomains, Fig. \ref{material_system}b. SAXS and calorimetry (Supplemental Material) show  $\mathrm{T_{I-N}(\approx 69 ^{\circ}C})$ and $\mathrm{T_{odt}(\approx 71 ^{\circ}C})$ are effectively coincident, indicating that self-assembly is driven by the isotropic-nematic transition on cooling, and that the system is weakly segregated. $\mathrm{T_{odt}}$ and $\mathrm{T_{N-SmA}\approx 64 ^{\circ}C}$ are visible in Fig. \ref{material_system}c.

Slow cooling (0.1 $^{\circ}$C/min.) from the disordered melt ($\mathrm{T>T_{odt}}$) to 25 $^{\circ}$C at 6 T results in pronounced alignment of the cylinders and LC director parallel to the field. Intense spot-like reflections from the cylindrical microdomains occur along the equatorial line with orthogonal reflections from the SmA layers, Fig. \ref{material_system}d. The progressive development of structural and orientational order on cooling is revealed by mapping the scattered intensity azimuthally, Fig. \ref{material_system}e. This depiction makes clear the coincidence of I-N and ODT, and further, a pronounced narrowing of the azimuthal intensity distribution for the BCP scattering on passage through N-SmA. This narrowing is shown clearly in the temperature dependence of the full-width at half maximum (FWHM) of Gaussian fits of the azimuthal intensity distributions, Fig. \ref{material_system}f (Supplemental Material).

Zero-field cooling experiments better elucidate the coupling between BCP and LC alignment at the N-SmA transition. Samples were first held at 6 T for 60 min. in the nematic at $\mathrm{68 ^{\circ}C}$ before cooling to $\mathrm{25 ^{\circ}C}$ under 0 T at various cooling rates. Data shown in Fig. \ref{alignment_data}a-b chart the development of the system in the vicinity of $\mathrm{T_{N-SmA}}$ when cooled at 0.1 $\mathrm{^{\circ}C/min}$. Orientation distribution coefficients  $\langle P_2\rangle =\langle \frac{1}{2}(3\cos^2\theta-1)\rangle$ are calculated using Gaussian fits of the SAXS azimuthal intensity for the BCP and SmA structures, $\mathrm{\langle P_2^{BCP}\rangle}$ and $\mathrm{\langle P_2^{SmA}\rangle}$, and for the 0.4-0.5 nm lateral mesogen ordering from WAXS data, $\mathrm{\langle P_2^{LC}\rangle}$. Here we are careful to distinguish orientation distribution coefficients from order parameters. Although they are similarly determined, they have different purposes. The former is simply a useful measure of the degree of alignment, $\mathrm{\langle P_2^{BCP}\rangle}$ and $\mathrm{\langle P_2^{SmA}\rangle}$, whereas the latter is a thermodynamically relevant descriptor. The thermodynamic order parameter in the nematic, $S$, is identical to the orientation distribution coefficient of the mesogens, $\mathrm{S=\langle P_2^{LC}\rangle}$.

The system shows a sudden and large increase of orientational order of the BCP microdomains and SmA layers, from $\langle P_2\rangle\sim0.55$ to nearly 1, in traversing the SmA transition, and a comparatively much smaller change of orientational order of the mesogens, Fig. \ref{alignment_data}c. The data highlight the efficacy of the molecular field developed in the nematic state in driving microdomain alignment in the absence of the magnetic field. The transformation occurs rapidly with alignment quality relatively insensitive to cooling rates as large as 10 $\mathrm{^{\circ}C/min}$, Fig. \ref{alignment_data}d-f.

Field-resolved experiments provide a complementary perspective to temperature-resolved data and also highlight the strong coupling between $\mathrm{\langle P_2^{SmA}\rangle}$ and $\mathrm{\langle P_2^{BCP}\rangle}$. Data are shown in Fig. \ref{alignment_data_cartoon}a-b for samples subjected to zero-field cooling after initial exposure in the nematic to fields of varying intensity. FWHM data show unambiguously that the change in alignment quality of the BCP is most strongly coupled to that of the SmA layers, and that changes in the alignment of the mesogens are minimal between the nematic and smectic states.

The N-SmA transition in prototypical small-molecule LCs is highlighted by the emergence of translational order, a 1-D density wave correlating the centers of mass of the smectic layers, Eq. \ref{eq:smectic_density}, with layer spacing $d$ and position $z$. The degree of translational order is captured by the Fourier coefficients $\sigma_n$ with the translational order parameter, $\Sigma$, taken as $\Sigma=\sigma_1$. The coupling of the N and SmA order parameters in part drives the complexity and weakly first-order nature of the N-SmA transition, with layering in the smectic suppressing director fluctuations \cite{mcmillan1971simple,meyer1976mean,Gramsbergen1988,anisimov1990experimental,demus1999physical}. Further, the elastic constants associated with twist ($K_{22}$) and bend ($K_{33}$) modes increase rapidly on nearing $\mathrm{T_{N-SmA}}$ \cite{karat1977elasticity,demus1999physical}.

\begin{equation}
\rho(z)=\langle\rho\rangle+\sum\limits_n \sigma_n\cos(2\pi nz/d)
\label{eq:smectic_density}
\end{equation}

As the orientational order of smectic layers is intrinsically coupled to the emergence of translational order, we speculate that the BCP orientational order is in fact due to, and should therefore be correlated with, the underlying 1-D smectic order. We hypothesize that this coupling is driven by microdomain persistence necessitated by emerging translational order in the SmA matrix and physical contiguity of the system. The concept is illustrated in Fig. \ref{alignment_data_cartoon}c. Stubby microdomains in the nematic couple to the local director field $\mathbf{\hat{n}_{loc}}$ through the mesogen planar anchoring condition. Variation of $\mathbf{\hat{n}_{loc}}$ as captured by $\mathrm{\langle P_2^{LC}\rangle}$ is present in $\mathrm{\langle P_2^{BCP}\rangle}$ in the nematic state. The translational order of the SmA drives an increase of aspect ratio of microdomains which enables coupling to the director as defined over a larger length scale, and therefore with less spatial fluctuation.

\begin{figure}[ht]
\centering
\includegraphics[width=80mm, scale=1]{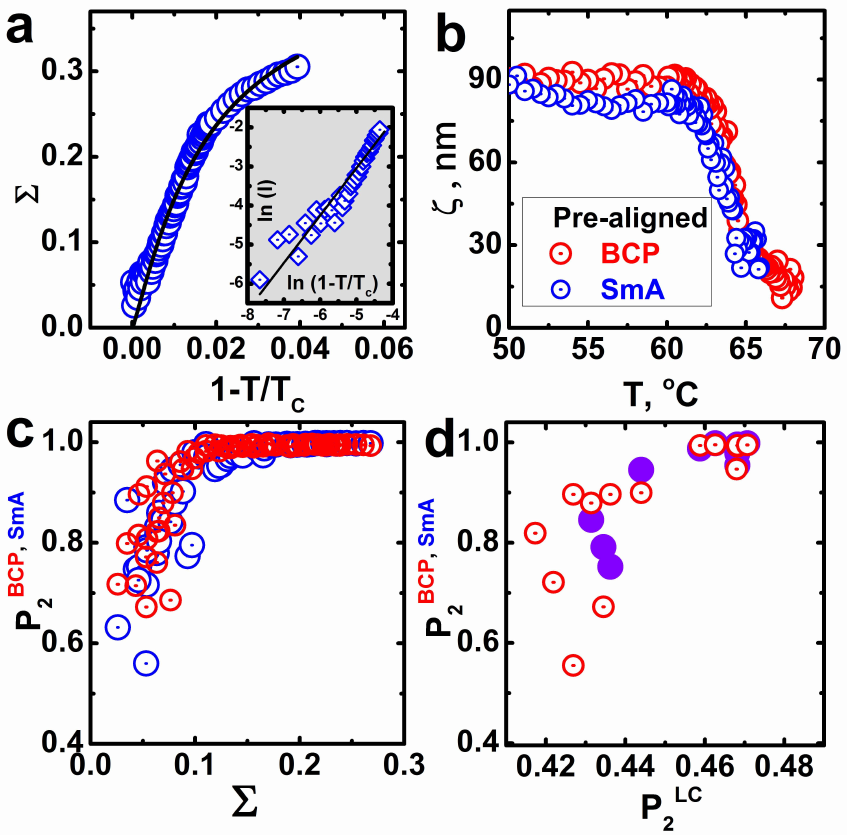}
\caption{a) SmA translational order parameter $\Sigma(T)$. Inset: Estimation of $I_0$ as $\ln I(T)=\ln I_0+2\beta\ln[1-(T/T_c)]$, Eq. \ref{eq:smectic_order_parameter}, with $\beta\approx 0.47$. b) Correlation lengths along the field direction for BCP and SmA structures. Instrument resolution limits prevent measurements of $\xi>90$ nm. c) and d) $\mathrm{\langle P_2^{BCP}\rangle}$ (solid symbols) and $\mathrm{\langle P_2^{SmA}\rangle}$ (open symbols) as functions of $\Sigma$ and $P_2^{LC}$ respectively reflecting the strong correlation of BCP alignment to the SmA translational ordering.}
\label{order_parameters}
\end{figure}

An examination of experimentally determined correlation lengths, $\xi$, and $\Sigma$ provides strong evidence for the above hypothesis. We determine $\Sigma$ from $I(T)$ of the smectic layers, using the approach of Giesselman \textit{et al.}, Eq. \ref{eq:smectic_order_parameter}, where $T_c$ is the upper stability limit that occurs in advance of $\mathrm{T_{N-SmA}}$ \cite{kapernaum2008simple}. $\xi$ is evaluated using the Scherrer formalism, $\xi=\lambda/\beta\cos\theta$ where $\beta$ is the integral peak breath, $2\theta$ the scattering angle, and $\lambda$ the x-ray wavelength. The intensity and correlation lengths of the BCP and SmA structures are tightly coupled, Fig. \ref{order_parameters}. Notably, the orientational order of the BCP is well correlated with the translational order parameter $\Sigma$ rather than the mesogen order parameter which changes only modestly in the N-SmA transition.

\begin{equation}
\Sigma^2=\frac{I(T)}{I_0}=\left[1-(T/T_c)\right]^{2\beta}
\label{eq:smectic_order_parameter}
\end{equation}

\begin{figure}
\centering
\includegraphics[width=85mm, scale=1]{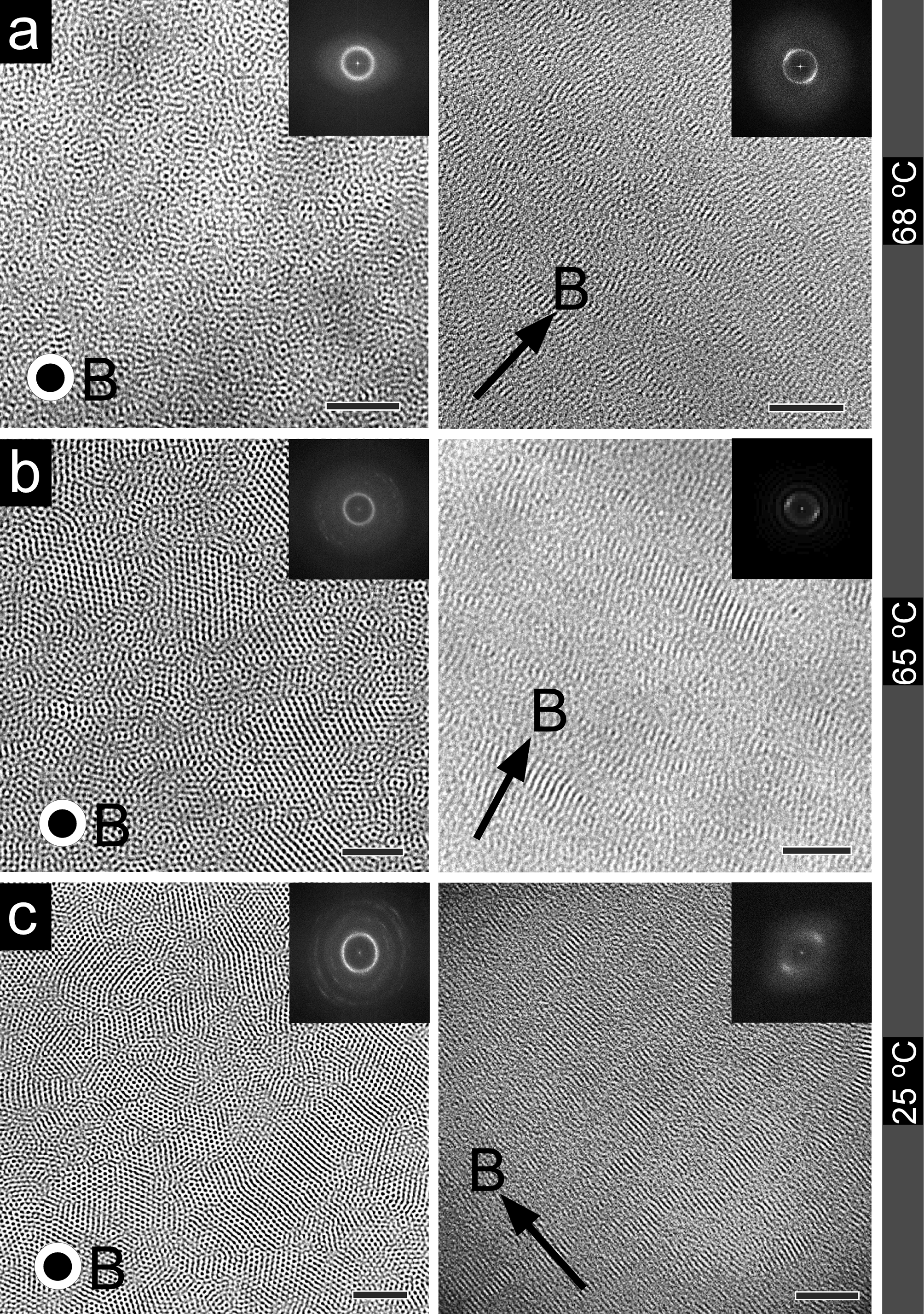}
\caption{TEM images along (left) and perpendicular to the field (right) for samples quenched from the nematic at 68 $^{\circ}$C (a), quenched from near $\mathrm{T_{N-SmA}}$ at 65 $^{\circ}$C (b), and in the SmA state at 25 $^{\circ}$C (c). PEO domains appear dark due to staining. Scale bar: 100 nm. FFTs are shown as insets.}
\label{tem}
\end{figure}

TEM data provide a complementary view. Samples quenched in liquid nitrogen and imaged at room temperature reveal the morphology at the relevant temperatures in the N and SmA states. TEM images reveal a poorly ordered nematic with microdomains of such low aspect ratio differentiating the morphology along versus perpendicular to the field direction is difficult, Fig. \ref{tem}a. The microdomains increase in aspect ratio significantly on entering the SmA, Fig. \ref{tem}b. Finally at room temperature, Fig. \ref{tem}c, distinct grains can be distinguished when viewed along the field direction. The fact that the increase in microdomain aspect ratio is driven by the N-SmA transition and does not occur simply as a consequence of the underlying temperature dependent interaction between the blocks is highlighted by a secondary system which displays only a nematic state, with $\mathrm{T_{NI}=45 ^{\circ}C}$ (Supplemental Material).

We identify a common thread connecting the present work with observations in systems involving nematic solvents. Onsager correctly anticipated that long rods would exhibit greater orientational order than shorter rods of the same diameter in polydisperse hard rod suspensions \cite{onsager1949effects}, as explored by Lekkerkerker \textit{et al} \cite{lekkerkerker1984isotropic}. Yodh \textit{et al.} considered polymer chain conformations in a nematic suspension of anisotropic colloids, \textit{fd} virus particles \cite{dogic2004elongation}. The polymer persistence length $l_p$ determined the occurrence of coil-rod transitions driven by the nematic field of the background fluid and the polymer orientational order in the rod-like state exceeded that of the nematic background. The disparity in orientational order is broadly consistent with Onsager's prediction as $l_p$ exceeds the length of the \textit{fd} nematogens. Here we observe similarly that $\mathrm{P_2^{BCP}}$ exceeds $\mathrm{P_2^{LC}}$ of the nematic matrix, Fig. \ref {alignment_data}c. While the analogy to the Onsager scenario is inexact, it is nonetheless useful in capturing the salient point that the orientational order of anisotropic objects in a nematic field increases with the length of the objects.

In conclusion, this work provides a detailed examination of the roles of the I-N and N-SmA transitions in dictating magnetic field-driven alignment of BCP microdomains embedded in the LC matrix. The I-N transition drives self-assembly of the weakly segregated blocks, resulting in a poorly-ordered and weakly-aligned fluid of low aspect ratio cylindrical microdomains in a nematic continuum under the field. SAXS and TEM point to an increase in the microdomain aspect ratio as the driver for the strong alignment of the system by what is in essence a molecular field (provided by the LC director) during zero field cooling experiments from the poorly ordered nematic. The N-SmA transition plays an important role in this process by providing the impetus for the extension of the cylindrical microdomains, reflecting the physical contiguity of the microdomains over larger length scales that is necessitated by the emerging SmA translational order. Additionally, the diverging elasticities $K_{22}$ and $K_{33}$ on approaching N-SmA may also contribute to $\mathrm{\langle P_2^{BCP}\rangle}$ by suppressing microdomain arrangements involving twist and bend deformation of the director field. Such a `smectic enhancement' of orientational order would be more relevant at smaller aspect ratios.

Anecdotal evidence has suggested that the existence of smectic order aids in magnetic field DSA of weakly segregated BCPs \cite{deshmukh2014molecular,gopinadhan2013order,gopinadhan2012magnetic}. The results here provide crucial insight as to why this is the case. The transition from the disordered melt sequentially via a high mobility but poorly ordered nematic, and a low mobility but better-ordered smectic, may provide a practical benefit in optimizing field response kinetics.


\begin{acknowledgments}
This work was supported by NSF under DMR-1410568 and DMR-0847534. Facilities use was supported by YINQE and DMR-1119826. The authors thank Dr. Pawe{\l} Majewski and Julie Kornfield for fruitful discussions, Mike Degen (Rigaku Inc.) and AMI Inc. for technical support.
\end{acknowledgments}

\bibliographystyle{apsrev}
\bibliography{smectic_refinement}

\section*{Supplemental Material}

\begin{figure}[h]
  \centering
    \includegraphics[width=85mm]{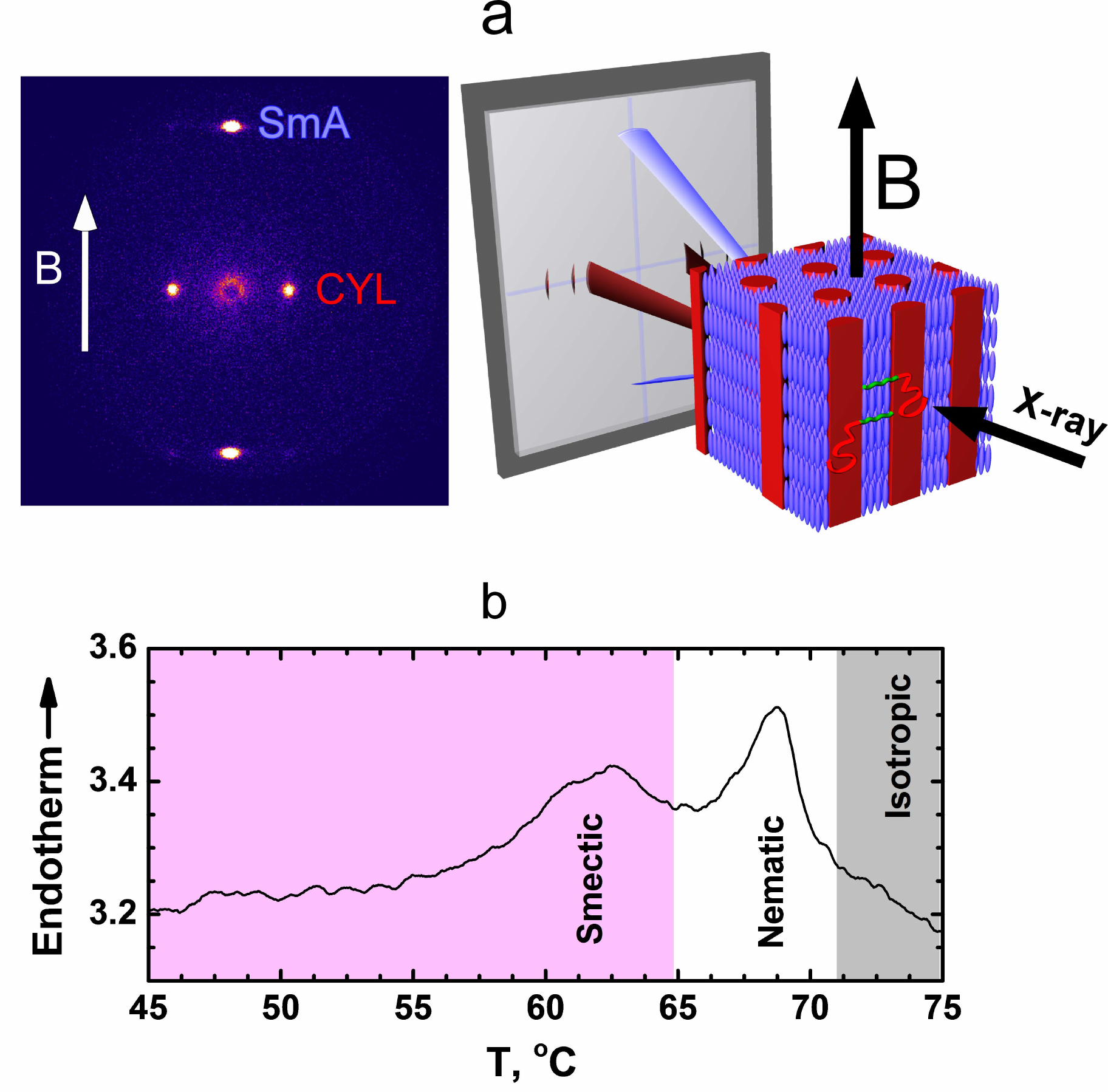}
    \caption{a) Schematic illustration of the scattering geometry for the \textit{in situ} SAXS measurements in the magnetic field (right), and 2-D SAXS data for comparison (left). b) Differential scanning calorimetry (DSC) data collected at a cooling rate of 1 $^{\circ}$C/min indicate $\mathrm{T_{N-SmA}\approx 64 ^{\circ}C}$ and $\mathrm{T_{I-N}\approx 69 ^{\circ}C}$.}
\end{figure}

\begin{figure}[h]
  \centering
    \includegraphics[width=85mm]{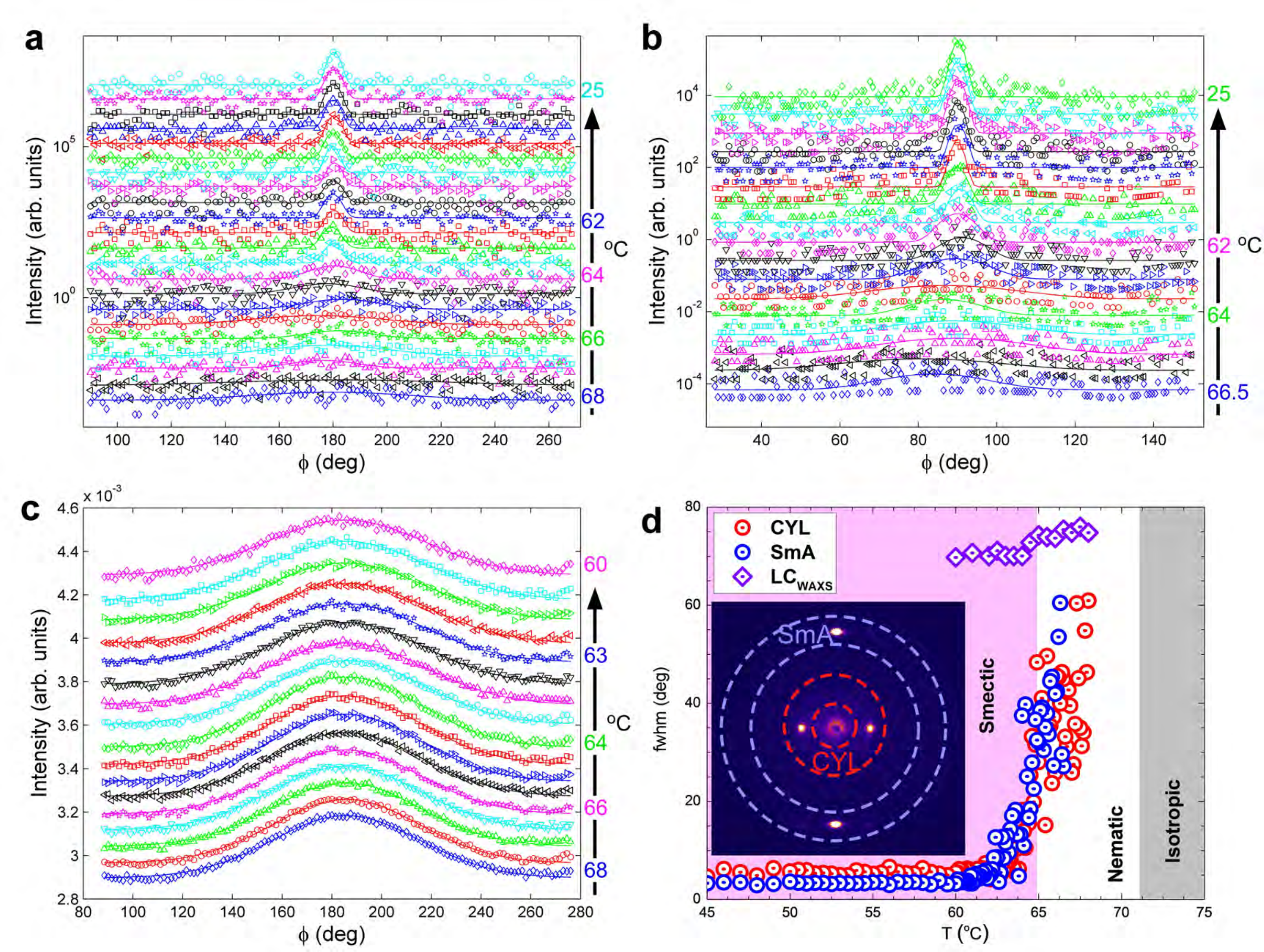}
    \caption{(a-c) Azimuthal intensity plots of scattering from primary peak of cylindrical microdomains, SmA structure and LC lateral correlations (WAXS regime), and the corresponding fwhm of the scattering peaks respectively (d). The sample was field-aligned in the nematic state at 68 $^{\circ}$C prior to zero field cooling to the smectic phase. The solid lines are Gaussian fit to the data.}
\end{figure}

\begin{figure}[h]
  \centering
    \includegraphics[width=85mm]{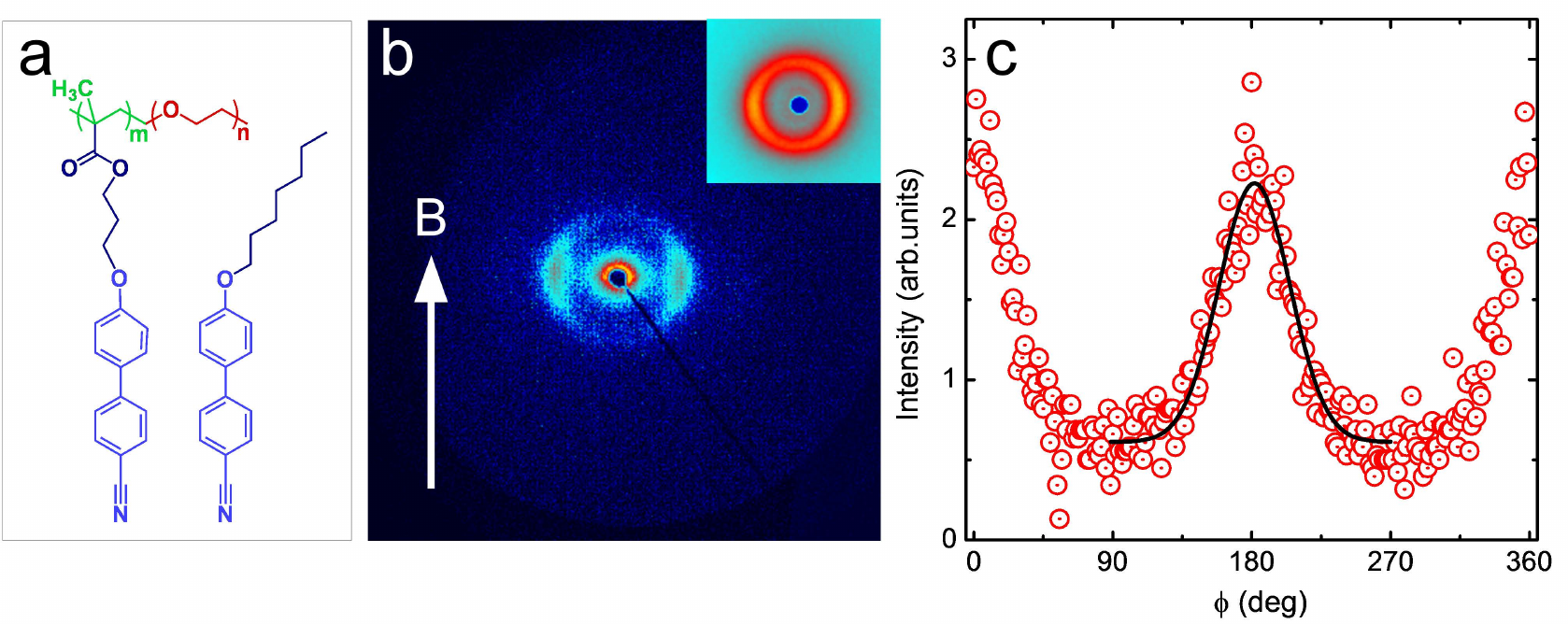}
    \caption{a) PEO-b-MA/LC system with 3 carbon spacer is blended with 6OCB (hexyloxycyanobipheny) free mesogens at a stoichiometry (molar ratio) R=0.75 of 6OCB MA/LC monomer units to obtain a nematic system with $\mathrm{T_{I-N}\approx 45 ^{\circ}}$C. b) The system is magnetically aligned at 6 T, and the resulting 2-D SAXS and WAXS (inset) are shown. c) Azimuthal intensity plot of the microdomain scattering. The solid line is the Gaussian fit to the data, which gives a fwhm of 54 degrees and orientation distribution coefficient of 0.63.}
\end{figure}

\end{document}